\documentclass[12pt,preprint]{aastex}
\begin{document}

\parindent=1.0cm

\title {YOUNG STARS AT THE EDGE: STELLAR CLUSTERING IN THE OUTER REGIONS OF THE M33 DISK \altaffilmark{1}}

\author{T. J. Davidge}

\affil{Herzberg Institute of Astrophysics,
\\National Research Council of Canada, 5071 West Saanich Road,
\\Victoria, BC Canada V9E 2E7\\ {\it email: tim.davidge@nrc.ca}}

\author{T. H. Puzia}

\affil{Department of Astronomy and Astrophysics,
\\Pontifica Universidad Catolica, Santiago, Chile 7820436
\\ {\it email: tpuzia@gmail.com}}

\author{A. W. McConnachie}

\affil{Herzberg Institute of Astrophysics,
\\National Research Council of Canada, 5071 West Saanich Road,
\\Victoria, BC Canada V9E 2E7\\ {\it email: alan.mcconachie@nrc.ca}}

\altaffiltext{1}{Based on observations obtained with MegaPrime/MegaCam, a joint 
project of CFHT and CEA/DAPNIA, at the Canada-France-Hawaii Telescope (CFHT) 
which is operated by the National Research Council (NRC) of Canada, the 
Institut National des Science de l'Univers of the Centre National de la 
Recherche Scientifique (CNRS) of France, and the University of Hawaii.}

\begin{abstract}

	We investigate the distribution of bright main sequence stars near the 
northern edge of the M33 disk. Clustering on sub-kpc scales is seen among 
stars with ages $\sim 10$ Myr, and two large star-forming complexes are 
identified. Similar large-scale grouping is not evident among stars 
with ages 100 Myr. These stars are also 
distributed over a much larger area than those with younger ages, 
and it is argued that random stellar motions alone, as opposed to orderly 
motions of the type spurred by large scale secular effects, 
can re-distribute stars out to distances of at least 2 kpc (i.e. one disk 
scale length) from their birth places on 100 Myr timescales. Such random motions 
may thus play a significant role in populating the outer regions 
of the M33 disk. Finally, it is suggested that --
to the extent that the ambient properties of the outer disk mirror those in 
the main body of the disk -- stars in this part of M33 
may have formed in star clusters with masses 50 -- 250 M$_\odot$, which is 
substantially lower than the peak of the solar neighborhood initial cluster 
mass function. 

\end{abstract}


\section{INTRODUCTION}

	The stellar disks of spiral galaxies can extend to many scale lengths 
(e.g. Davidge 2006; Pohlen \& Trujillo 2006), and the stars that populate the 
peripheral regions of disks likely have a range of origins. Some of the 
stars at large radii probably formed {\it in situ}. Ultraviolet light 
concentrations that trace young stellar regions are seen at large radii in some 
nearby spiral galaxies (e.g. Gil de Paz et al. 2008; Zaritsky \& Christlein 
2007). While the density of interstellar material at large radii tends to be 
too low to trigger large-scale star formation, localized density enhancements 
may occur as a result of compression from spiral density waves (e.g. Bush et al. 
2008). The presence of a dark baryonic component in the disk plane could also 
enable star formation in areas where the gas density may otherwise appear to 
be too low (Revaz et al. 2009). Some fraction of the stars in the outer disk 
are probably migrants from smaller radii, as recent studies have shown that 
secular processes (e.g. Roskar et al. 2008) and radial mixing induced by 
interactions (e.g. Quillen et al. 2009) can contribute significantly to the 
stellar contents of the outer regions of disks.

	The distribution of stars in the nearest spiral galaxies 
provide clues into the processes that populate the outermost regions of disks.
Star-forming activity at large radii produces distinct signatures in galactic 
light and color profiles (e.g. Sanchez-Blazquez et al. 2009). As for secular 
effects, the processes that re-distribute stars throughout disks act in a 
cumulative manner on stellar orbits, with the result that stars that have 
moved the furthest from their places of birth will tend to be the oldest 
-- stars with progressively older ages may thus be found at progressively larger 
galactocentric distances (e.g. Roskar et al. 2008), in contradiction to 
what might niavely be expected due to inside-out disk formation. 

	The present letter is part of a larger study of young and intermediate 
age stars throughout the disk of M33 (Davidge et al. 2011, in 
preparation). The entire dataset consists of five MegaCam pointings, 
and here we examine the distribution of stars in an area that includes two 
of the most remote star-forming complexes in M33. 
A distance modulus of 24.93 (Bonanos et al. 2006) is adopted. 
Recent distance modulus estimates for M33 show a spread of a 
few tenths of a dex, and the Bonanos et al. value was selected because it is 
based on eclipsing binaries, which are a primary distance indicator.
 
\section{OBSERVATIONS, REDUCTIONS, \& PHOTOMETRIC MEASUREMENTS}

	The data were recorded as part of the 2009B 
MegaCam (Boulade et al. 2003) observing queue on the Canada-France-Hawaii 
Telescope (CFHT). The MegaCam detector is a mosaic of 36 
$2048 \times 4612$ E2V CCDs that are deployed in a $4 \times 9$ format. A 
$0.96 \times 0.94$ degree$^2$ area is imaged with 0.18 arcsec pixel$^{-1}$ 
sampling. Five 150 sec exposures were recorded in $g'$, and ten 440 sec 
exposures were obtained in $u'$. 

	The initial processing of the data, which included 
bias subtraction and flat-fielding, was done with the ELIXER 
pipeline at the CFHT. The ELIXER-processed images were aligned, 
stacked, and then trimmed to the area of common exposure time.
This paper deals with objects in a single $18 \times 13.5$ arcmin$^2$ 
field that is centered at 01:34:43 Right Ascension and 31:22:00 
Declination (E2000), and samples the disk of M33 at a galactocentric distance 
of 8.3 kpc, or 4 disk scale lengths. The southern 
half of this area, which includes the northern spiral 
arm and two young stellar complexes, is shown in Figure 1. 
Stars in the final images have FWHM $\sim 0.9$ arcsec.

	The brightnesses of individual stars were measured 
with the PSF-fitting routine ALLSTAR (Stetson \& Harris 1988). 
The photometric calibration was defined using zeropoints and transformation 
coefficients computed from standard star observations that were recorded 
during 2009B. Sources that depart from the trend between magnitude and 
the photometric error computed by ALLSTAR, which 
tend to be non-stellar in appearance (e.g. Davidge 2010), 
were removed from the photometric catalogue. 

\section{RESULTS}

	The $(u', g'-u')$ CMDs of stars in the `Spiral Arm' and `Outer Disk' 
areas indicated in Figure 1 are shown in Figure 2. The `Spiral Arm' region 
covers the diffuse distribution of stars in the northern spiral 
arm, while the `Outer Disk' area covers the remainder of the field. There is a 
large number of bright main sequence stars in the Spiral Arm CMD, and a
comparison with Z = 0.004 isochrones from Girardi et al. (2004) indicates that 
stars with ages from $\leq 10$ Myr to $\geq 100$ Myr are detected. 
This metallicity was selected based on the oxygen 
abundance seen throughout much of the M33 disk (e.g. Magrini et al. 2010), 
although the predicted locus of the upper main sequence is not 
sensitive to the adopted metallicity. The main sequence in the Outer Disk CMD is 
much less pronounced than in the Spiral Arm CMD; still, that modest 
numbers of young and intermediate age main sequence stars are present 
indicates that a diffusely distributed, young stellar component occurs outside 
of the main body of the spiral arm.

	The locations on the sky of stars in three 
areas of the $(u', g'-u')$ CMD, marked in Figure 2, are shown in Figure 
1. There are obvious age-related differences, in the sense that 
the stars in the 10 Myr sample tend to group together more than 
those in the 40 and 100 Myr samples. Even though the stellar 
distribution becomes more diffuse towards older ages, 
the northern spiral arm can still be identified in the 
100 Myr sample. While not shown here, the overall distribution of 
older samples (e.g. those with $u'$ between 24.5 and 25.5, and $u'-g'$ 
between 0 and 1, which have an age $\sim 200$ Myr), is even more diffuse. 

	The extent of clustering \footnote[2]{In this paper, clustering 
refers to the grouping of stars over a range of spatial scales. This 
includes, but is not restricted to, objects that are in 
`star clusters', which typically subtend only a few parsecs, and so are not 
resolved with these data.} can be quantified by examining the angular 
separations between star -- star pairs, and we refer to the histogram 
distribution of all possible pairings as the star--star separation function 
(S3F). The S3F is based on a simple observable -- an angular measurement on the 
sky -- and yields information about the large scale distribution of objects 
that can be difficult to quantify by eye. The gaps between the CCDs do not 
significantly affect the S3Fs, as these amount to only $\sim 4\%$ 
of the field covered.

	We first consider the S3F of sources in the northern half of the 
field. This area is well offset from the disk, and contains 
a mix of halo stars and unresolved galaxies, but few -- if any -- young 
or intermediate age stars belonging to M33. Thus, this area serves as a 
control for investigating the distribution of objects at smaller radii.
The S3F of objects in the northern half of the field that fall 
within the 100 Myr region of the CMD is shown in the bottom panel of 
Figure 3. The gradual drop-off in the separation frequency 
at separations $> 450 -- 500$ arcsec is due to the finite size of the area 
sampled. The S3F is not symmetric because the field is not square. 
The inflexion point of the S3F of uniformly distributed objects will occur at a 
scale that is roughly one half the length of the shortest axis of the area 
examined, which is 425 arcsec (2 kpc), and this matches approximately the 
inflexion point in the bottom panel of Figure 3. 

	The S3Fs of main sequence stars in the southern half of the MegaCam 
sub-panel are shown in the top three panels of Figure 3. These 
S3Fs clearly differ from the S3F of sources 
in the northern half of the MegaCam data, due to stellar grouping in M33. 
The S3F of the 10 Myr sample contains substantial signal at separations 
$r < 250$ arcsec ($< 1.1$ kpc), and the width of the peak at small 
separations indicates that the youngest stars are grouped on scales $< 150$ 
arcsec ($< 700$ pc). This is comparable to the dimensions of 
star-forming complexes in nearby galaxies (e.g. Efremov 1995), as well as 
the scale of coherent star-formation in M33 (Sanchez et al. 
2010). There is a second peak in the 10 Myr S3F near $r \sim 450$ (2.1 kpc) 
arcsec, and this occurs because the two young stellar concentrations in this 
field beat against each other in the separation measurements; the separation 
between the two clumps in the upper right hand panel agrees with that between 
the two most prominent clumps in Figure 1.

	When compared with the 10 Myr S3F, clustering signatures are 
broader and have a smaller amplitude in the 
S3F of the 40 Myr sample. The majority of stars in the 
40 Myr sample are separated by distances up to at least 300 arcsec (1.4 kpc). 
Even at this comparatively young age, stars have moved distances that 
are large enough to significantly blur clustering signatures in the S3F.

	There is a more-or-less uniform signal in the 100 
Myr S3F between 150 and 400 arcsec (0.7 and 1.9 kpc), with no evidence of 
clustering at separations $< 150$ arcsec ($< 0.7$ kpc) -- large stellar 
complexes evidently dissipate over $\sim 100$ Myr timescales in this part of 
M33. In fact, the width of the ramp-up in the 100 Myr S3F 
suggests that the minumum star-star separation is typically $\sim 50$ arcsec, 
or $\sim 0.25$ kpc, for stars of this age, while the onset of the plateau 
in the S3F suggests that the typical star-star separation is at least 150 
arcsec, or $\sim 0.7$ kpc.

\section{DISCUSSION \& SUMMARY}

	The star-star separation function (S3F) has been used to investigate 
the projected distribution of main sequence stars in the northern disk of 
M33. Two young stellar complexes produce significant 
signal in the S3F of stars with ages $\sim 10$ Myr 
at separations $r < 150$ arcsec ($d < 0.7$ kpc). However, signatures of 
clustering are greatly diminished among stars with ages $\sim 40$ Myr, 
and the smooth S3F of stars with ages $\sim 100$ Myr suggests that there is 
little if any large-scale clustering among these stars. Thus, large scale 
stellar structures in this part of M33 evidently dissipate over time scales 
$\leq 100$ Myr.

	Stellar complexes in the outer regions of disks may be subjected 
to disruption mechanisms that differ from those in the main body of the disk. 
There is evidence for heating by halo structures in the outer regions of 
nearby galaxies (e.g. Martin \& Kennicutt 2001), and dynamical measurements 
suggest that halo bombardment becomes a significant source of heating 
at 4 disk scale lengths in nearby spirals (Herrmann et al. 
2009), and this is the part of the M33 disk that we examine here. 
The broad, evenly distributed signal in the 100 Myr S3F between 
150 and 450 arcsec (0.7 -- 2.1 kpc) results from random motions on the 
order of $\sim 20$ km sec$^{-1}$, and this is comparable to the outer disk 
extraplanar motions measured by Herrmann et al. (2009). Putman et al. 
(2009) find that HI in the outer regions of M33 has a velocity dispersion of 
18.5 km sec$^{-1}$, and suggest that this may be a relic of an interaction 
within the past few Gyr between M31 and M33. 

	Newly formed stellar systems will be more prone to disruption if 
they have a low star formation efficiency (SFE), as feedback will remove gas 
early-on, thereby reducing -- perhaps catastrophically -- the gravitational 
field of the nascent system (Lada \& Lada 2003). A general trend for the SFE to 
diminish towards larger radii is seen in nearby galaxies (Leroy et al. 2008). 
This result is based on measurements made over kpc spatial scales, which is 
comparable to the sizes of the large structures probed here,

	Star clusters are sub-structures within the large-scale complexes that 
are investigated here. The largest disk star clusters in M33 subtend $\leq 2$ 
arcsec (San Roman et al. 2010), and so fall in the smallest bin in the S3F. 
The signal in the S3F of the 100 Myr sample in the 0 - 20 arcsec bin is markedly 
smaller than in the 10 Myr sample, and if this trend extends to sub-arcsec 
sizes then this will be consistent with stellar clusters dissipating over 
$\sim 0.1$ Gyr timescales. In fact, the spatial distribution 
of star clusters with ages $< 0.1 - 0.3$ Gyr in M33 is more 
compact than that of stars with the same age (Sarajedini \& Mancone 2007; 
Roman et al. 2010), suggesting that young star clusters in M33 dissipate 
over time spans that are less than a few tenths of a Gyr. 

	The disruption timescale of star clusters 
depends on a number of factors, including the rate at which remnant gas is 
removed from the cluster, the local environment, and two-body relaxation (e.g. 
summary by Elmegreen \& Hunter 2010). Gratier (2010) find that the masses of 
molecular clouds decrease with increasing radius in M33, and this should 
result in lower star cluster masses, which in turn may lead to a comparatively 
rapid disruption timescale for clusters in the outer regions of M33. 
Lamers et al. (2005a) estimates that a $10^4$ M$_{\odot}$ cluster 
in the main body of M33 typically disrupts after $\sim 1$ Gyr. Assuming no 
radial changes in the sources of dynamical heating, the ambient mass mixture 
that dominates the gravitational field, and the mean SFE 
within M33, then if the cluster disruption timescale $\propto$ 
mass$^\gamma$, where $\gamma =$ 0.62 (Baumgardt \& Makino 2003; Lamers et al. 
2005b), then the majority of stars in the outer disk of M33 formed in clusters 
with masses $\leq 50 - 250$ M$_{\odot}$ if they are disrupted on timescales 
of $\sim 100$ Myr. This characteristic cluster mass is roughly two orders of 
magnitude lower than the peak of the solar neighborhood initial cluster mass 
function predicted by Parmentier et al. (2008) and Kroupa \& Boily (2002). 
In fact, this is an upper limit to the initial cluster 
mass, in the sense that the pace with which clusters 
dissolve depends on factors such as the local mass density and the initial 
cluster mass, and a 10$^4$ M$_{\odot}$ cluster in the peripheral regions of the 
M33 disk would be even longer lived than predicted by Lamers et al. (2005a). 
Thus, if star clusters are disrupted over $\sim 0.1$ Gyr timescales 
in the outer regions of M33 then we predict that the star clusters found 
there will have (1) young ages, and (2) lower masses than those at smaller radii.

	We close by noting that the orderly large scale 
motions induced by secular processes are probably not significant among stars 
of the age considered here, given that the rotation period of 
the M33 disk is 200 - 300 Myr (Corbelli \& Salucci 2000). Rather, 
stars with ages $\sim 100$ Myr in this part of M33 appear to have obtained 
random stellar motions that allow them to populate regions up to $\sim 2$ kpc 
from where they formed. This effectively pushes out the observational 
boundary of the young disk.

\clearpage

\begin{figure}
\figurenum{1}
\epsscale{0.75}
\plotone{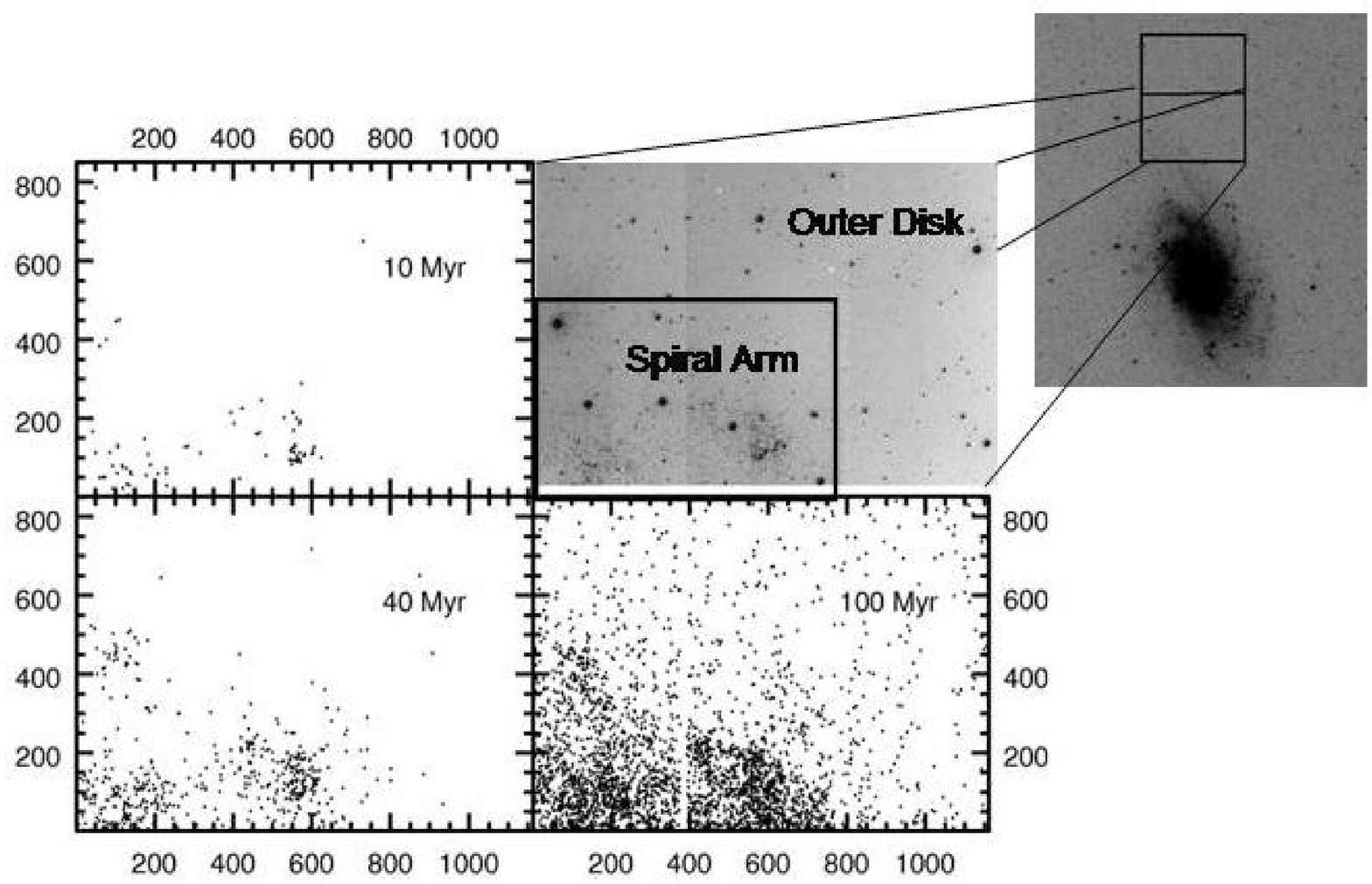}
\caption
{The distribution of objects in the area studied. 
North is at the top, and East is to the left. 
The field location is indicated on a $50 \times 60$ arcmin$^2$ section of 
the blue DSS; the line in this inset divides the field into the northern 
and southern halfs that are discussed in \S 3. 
A $18 \times 13.5$ arcmin$^2$ area is covered in each of the remaining 
panels, with the axes indicating distances in arcsec as measured from the south 
east corner. The areas containing the stars that are used to construct the CMDs
in Figure 2 are indicated on the $u'$ image. The locations of 
objects in three photometrically-selected samples are shown in the 
other panels. Note that (1) density enhancements due to two large young 
stellar complexes are seen in the 10 and 40 Myr samples, and (2) significant 
numbers of stars with ages $\sim 100$ Myr are seen to the north west of the 
main body of the spiral arm.}
\end{figure}

\clearpage

\begin{figure}
\figurenum{2}
\epsscale{0.75}
\plotone{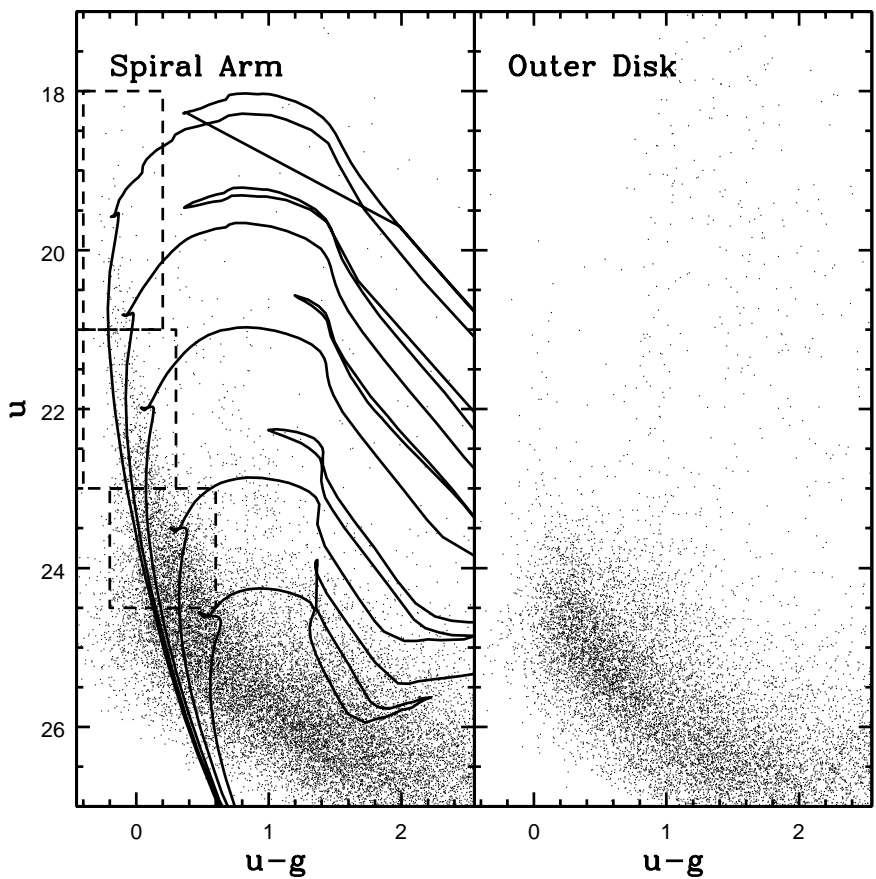}
\caption
{The $(u', u'-g')$ CMDs of sources in the `Spiral Arm' and `Outer Disk' 
regions indicated in Figure 1. The solid lines are Z = 0.004 isochrones from 
Girardi et al. (2004) with ages 10 Myr, 20 Myr, 40 Myr, 100 Myr, and 200 Myr. 
A foreground extinction A$_B = 0.181$ (Schlegel et al 1998) and an internal 
extinction A$_B = 0.16$ (Pierce \& Tully 1992) have been adopted. The dashed 
lines mark the areas on the CMDs from which stars are extracted to investigate 
clustering. MSTO stars near the centers of these boxes have ages 10 Myr, 
40 Myr, and 100 Myr.}
\end{figure}

\clearpage

\begin{figure}
\figurenum{3}
\epsscale{0.75}
\plotone{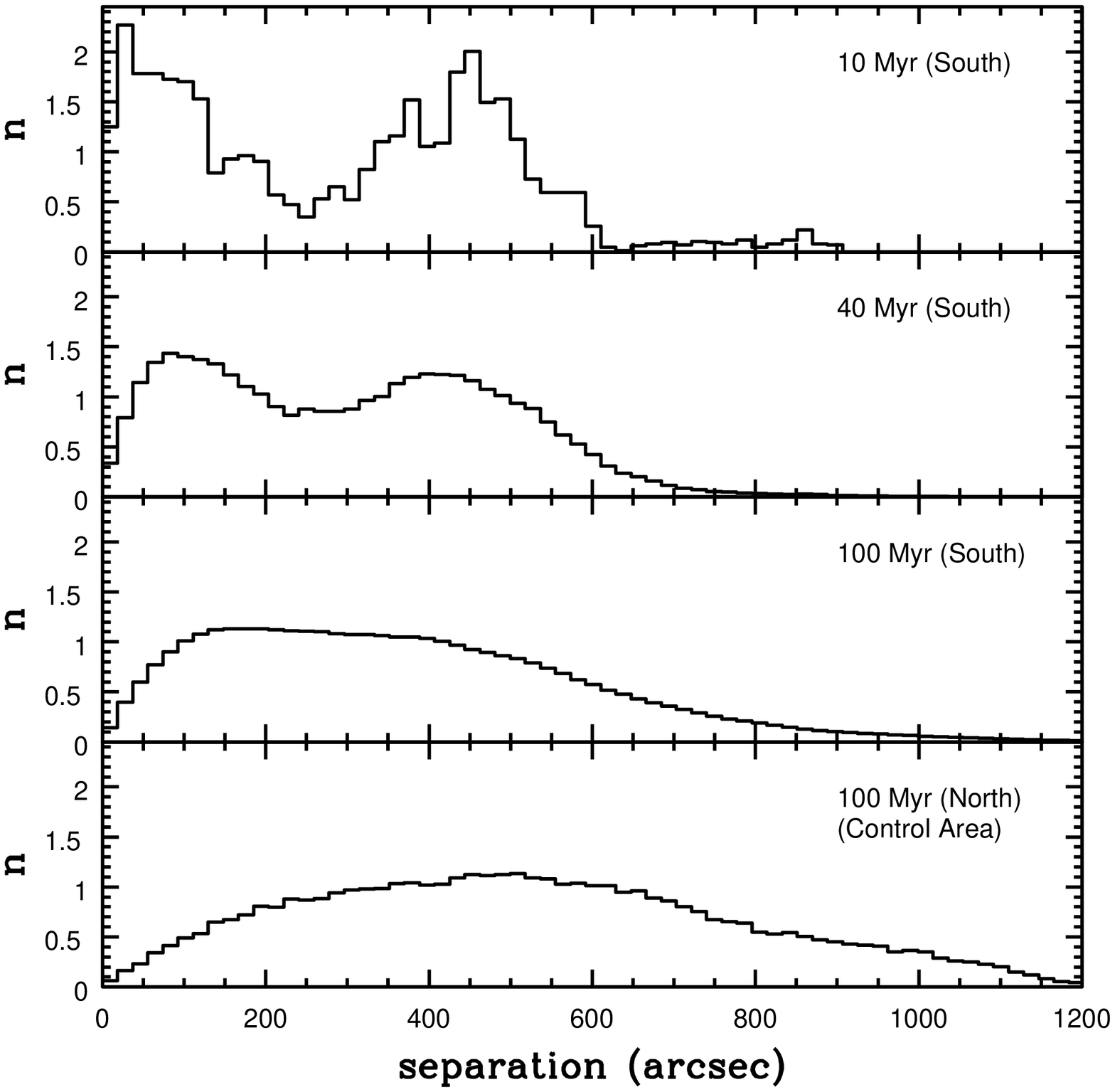}
\caption
{The star--star separation functions (S3Fs) of sources in the 
northern and southern half of the MegaCam data. 20 arcsec 
($\sim 90$ pc) binning has been used, and `n' is the number of sources 
in each bin normalized to the mean signal between 200 -- 550 arcsec, which 
is the range of separations that dominate disk stars in this field. 
The spatial scales investigated here probe large-scale stellar complexes, 
rather than classical star clusters, the majority of which in M33 tend to have 
angular sizes that are comparable to the seeing disk. 
The 100 Myr S3F of the northern half of the field is that of a more-or-less 
uniformly distributed ensemble of objects. In contrast, the S3Fs of sources 
in the southern half of the field show enhanced signal at separations 
$\leq 400$ arcsec. The prominent peak in the 10 Myr S3F at low separations 
indicates that the youngest stars group over angular scales $\leq 
150$ arsec, or projected distances of roughly 700 pc; the second peak is 
due to sources in the two complexes beating against each other. Aside from 
the broad signature produced by the main body of the northern spiral arm, there 
is no evidence of structures in the 100 Myr S3F, indicating that the 
processes that dissolve large scale star-forming regions in this part of M33 
do so on times scales $\leq 100$ Myr.}
\end{figure}


\parindent=0.0cm

\begin{references}

\reference{}Baumgardt, H., \& Makino, J. 2003, MNRAS, 340, 227

\reference{}Bonanos, A. Z., et al. 2006, ApJ, 652, 313

\reference{}Boulade, O., et al. 2003, Proc. SPIE, 4841, 72

\reference{}Bush, S. J., Cox, C. J., Hernquist, L., Thilker, D., \& Younger, J. D. 2008, ApJ, 683, L13

\reference{}Corbelli, E., \& Salucci, P. 2000, MNRAS, 311, 441

\reference{}Davidge, T. J. 2006, ApJ, 641, 822

\reference{}Davidge, T. J. 2010, ApJ, 725, 1342

\reference{}Dong, H., Calzetti, D., Regan, M., Thilker, D., Bianchi, L., Meurer, G. R., \& Walter, F. 2008, AJ, 136, 479

\reference{}Efremov, Y. N. 1995, AJ, 110, 2757

\reference{}Elmegreen, B. G., \& Hunter, D. A. 2010, ApJ, 712, 604

\reference{}Gil de Paz, A., et al. 2008, Formation and Evolution of Galaxy Disks, ASP Conf. Vol. 396, ed J. G. Funes \& E. M. Corsini, pp. 197

\reference{}Girardi, L., Grebel, E. K., Odenkirchen, M., \& Chiosi, C. 2004, A\&A, 422, 205

\reference{}Gratier, P. 2010, PhD thesis, Universite Bordeaux 1, in preparation

\reference{}Herrmann, K. A., Ciardullo, R., \& Sigurdsson, S. 2009, ApJ, 693, L19

\reference{}Kroupa, P., \& Boily, C. M. 2002, MNRAS, 336, 1188

\reference{}Lada, C. J., \& Lada, E. A. 2003, ARA\&A, 41, 57

\reference{}Lamers, H. J. G. L. M., Gieles, M., \& Portegies Zwart, S. F. 2005a, A\&A, 429, 173

\reference{}Lamers, H. J. G. L. M., Gieles, M., Bastian, N., Baumgardt, H., Kharchenko, N. V., \& Portigies Zwart, S. 2005b, A\&A, 441, 117

\reference{}Leroy, A. K., Walter, F., Brinks, E., Bigiel, F., de Blok, W. J. G., Madore, B., \& Thornley, M. D. 2008, AJ, 136, 2782

\reference{}Magrini, L., Stanghellini, L., Corbelli, E., Galli, D., \& Villaver, E. 2010, A\&A, 512, 63

\reference{}Martin, C. L., \& Kennicutt, R. C. Jr. 2001, ApJ, 555, 301

\reference{}Parmentier, G., Goodwin, S. P., Kroupa, P., \& Baumgardt, H. 2008, ApJ, 678, 347

\reference{}Pierce, M. J., \& Tully, R. B. 1992, ApJ, 387, 47

\reference{}Pohlen, M., \& Trujillo, I. 2006, A\&A, 454, 759

\reference{}Putnam, M. E., et al. 2009, ApJ, 703, 1486

\reference{}Quillen, A. C., Minchev, I., Bland-Hawthorn, J., \& Haywood, M. 2009, MNRAS, 397, 1599

\reference{}Revaz, Y., Pfenniger, D., Combes, F., \& Bournaud, F. 2009, A\&A, 501, 171

\reference{}San Roman, I., Sarajedini, A., \& Aparicio, A. 2010, ApJ, 720, 1674

\reference{}Sarajedini, A., \& Mancone, C. L. 2007, AJ, 134, 447

\reference{}Roskar, R., Debattista, V. P., Quinn, T. R., Stinson, G. S., \& Wadsley, J. 2008, ApJ, 684, L79

\reference{}Sanchez, N., Anez, N., Alfaro, E. J., \& Odekon, M. C. 2010, ApJ, 720, 541

\reference{}Sanchez-Blazquez, P., Courty, S., Gibson, B. K., \& Brook, C. B. 2009, MNRAS, 398, 591

\reference{}Schlegel, D. J., Finkbeiner, D. P., \& Davis, M. 1998, ApJ, 500, 525

\reference{}Stetson, P. B., \& Harris, W. E. 1988, AJ, 96, 909

\reference{}Zaritsky, D., \& Christlein, D. 2007, AJ, 134, 135

\end{references}
\end{document}